\documentstyle[12pt]{article}

\font\twelvemsb=msbm10 scaled 1200 
\def\Bbb#1{\hbox {\twelvemsb#1}}

\newcommand{\M}{{\cal M}}
\newcommand{\N}{{\cal N}}
\newcommand\bm[1]{\mbox{\boldmath$#1$}}
\newcommand\cm[1]{\bm{\cal#1}} 
\newcommand{\C}{\cm{C}} 
\newcommand\ov[1]{\overline{#1}} 
\newcommand{\F}{\bm{{\cal F}}}  
\newcommand{\FF}{\bm{F}}
\newcommand{\ssigma}{{\,\,\,\sigma}} 
\newcommand{\er}{\sigma}
\newcommand{\z}{\hat{Z}}

\newcommand{\I}{{\cm I}}
\newcommand{\Fsq}{\bm{{\cal F}^2}}
\newcommand{\f}{f}
\newtheorem{lemma}{Lemma}
\newtheorem{theorem}{Theorem}
\newtheorem{definition}{Definition}
\newtheorem{proposition}{Proposition}
\begin{document}

\title{A spacetime characterization of the Kerr metric} \author{Marc
Mars\thanks{Also at Laboratori de F\'{\i}sica Matem\`atica,
Societat Catalana de F\'{\i}sica, IEC, Barcelona,Spain.}\\  Institute of
Theoretical Physics, University of Vienna, \\ Boltzmanngasse 5, A-1090
Vienna, Austria.}  
\date{5 January 1999}
\maketitle
\begin{abstract}
We obtain a characterization of the Kerr metric among
stationary, asymptotically flat, vacuum spacetimes,
which extends the characterization in terms of the Simon tensor (defined only
in the manifold of trajectories) to the whole spacetime. 
More precisely, we define a three index tensor on any spacetime
with a Killing field, which vanishes identically for Kerr and which
coincides in the strictly stationary region with the Simon tensor when
projected down into the manifold of trajectories.
We prove that a stationary asymptotically flat vacuum 
spacetime with vanishing spacetime Simon tensor is locally isometric to Kerr.
A geometrical interpretation of this characterization in terms of the Weyl
tensor is also given. Namely,
a stationary, asymptotically flat vacuum spacetime such that each principal
null direction of the Killing form is a repeated principal null
direction of the Weyl tensor is locally isometric to Kerr.

\end{abstract}


\newpage

\section{Introduction.}

The Kerr metric plays a very prominent r\^{o}le in Einstein's theory
of general relativity. Obtained by R.Kerr in 1963 \cite{Kerr}, it 
was the first explicitly known stationary rotating (i.e. non-static)
asymptotically flat vacuum spacetime. Although many
other stationary and axially
symmetric asymptotically flat vacuum solutions
are known at present, the Kerr metric remains,
somehow, the simplest one.
More importantly, the Kerr spacetime
has a very special status among stationary vacuum solutions due to
the black hole uniqueness theorem, which states, roughly
speaking, that the exterior geometry of a stationary, asymptotically flat,
vacuum black hole must be Kerr. Despite its importance, the
reasons why this geometry plays such a privileged r\^{o}le remain
somewhat obscure. In comparison, the Schwarzschild metric has  nice
uniqueness
properties; in addition to Birkhoff's theorem, this
metric is well-known to
be the only static and asymptotically flat vacuum solution such that
the induced metric of the hypersurfaces orthogonal to the static
Killing vector are conformally flat. 

This leads one to consider whether there
exists any property of the Kerr metric which singles it out among
the class of stationary, asymptotically flat vacuum metrics. Some
characterizations are already known.
The first such was found by B. Carter \cite{Carter1} who
analyzed stationary and axisymmetric electrovacuum 
spacetimes such that the Hamilton-Jacobi and Schr\"odinger
equations are separable in some adapted coordinates and found a finite
parameter family of solutions containing the Kerr metric.
However, this analysis relies on coordinate
dependent conditions, which is not appropriate for characterizing
a spacetime. Furthermore, axial symmetry is assumed from the outset.
A second characterization
was obtained by Z. Perj\'{e}s \cite{Perjes1}, who restricted the analysis to
the strictly stationary subclass (i.e. to spacetimes with a Killing
vector which
is timelike everywhere). Working on the manifold of trajectories (i.e.
the set of orbits of the isometry group,
which is assumed to be a manifold),
Perj\'es showed that the Kerr metric can be found uniquely 
by demanding the existence of ``geodesic eigenrays'' of the Killing vector.
The third and, probably the  most interesting of the known
characterizations of Kerr is due to
W. Simon \cite{Simon}, who defined a tensor (now called Simon tensor)
on the manifold of trajectories which is identically zero
for the Kerr metric. 
Conversely, if the Simon tensor vanishes 
in an asymptotically flat, vacuum spacetime,
then there exists an
open neighbourhood of infinity which can be isometrically embedded
into the Kerr spacetime. Moreover, the Simon tensor
is equivalent to the Cotton tensor (which vanishes in
a three-dimensional Riemannian manifold if and only if
the metric is locally conformally flat) when the
Killing vector is hypersurface orthogonal,
so that one of the known characterizations
of Schwarzschild is recovered.
Despite the clear interest of this characterization
it still suffers  from two drawbacks.
The first one is that the theorem establishes the
isometry with Kerr only in some neighbourhood of infinity. 
Even though any strictly stationary vacuum metric is analytic
\cite{mzHaagen}, this does not
ensure that the isometry near infinity extends everywhere, because
analytic extensions of manifolds need not be unique.
However, this is probably not a serious objection 
and could presumably be fixed
by exploiting the interesting results by Z.Perj\'es in \cite{Perjes2},
who found all the possible local
forms of the metric in any strictly stationary spacetime with
vanishing Simon tensor. In particular, Perj\'es found that the most
general metric with those properties depends only on a few parameters,
thus showing that the asymptotically flatness condition in the
characterization of Kerr is only necessary in order to fix
the value of some constants.

The second objection is more serious and will be the object of
this paper. Since the whole construction of the Simon tensor
is done on the manifold of trajectories, 
the characterization in terms of the Simon tensor only works in the
strictly stationary region, i.e. outside any ergosphere.
However, for most practical
purposes (in particular, for the black hole uniqueness theorems)
the relevant region of Kerr is the
domain of outer communication, which contains a non-empty
ergoregion. Therefore the characterization in terms of objects
defined in the manifold of trajectories is not sufficient. Thus, it is
necessary to find a property of Kerr that distinguishes this
metric irrespectively of the norm of the Killing vector.
 
In order to treat this problem, we should look
for a characterization involving only spacetime
objects. Due to the relevance of the Simon tensor,
the obvious strategy is to try and
obtain a spacetime version of the Simon tensor in the
region where the Killing vector is timelike and analyze
whether or not this spacetime tensor remains regular
at points where the Killing vector becomes null. Provided this
is the case, it would remain to show that the extended Simon tensor
vanishes
everywhere for Kerr and prove also that the converse holds, i.e.
that an asymptotically flat, vacuum spacetime with vanishing
spacetime Simon tensor is locally isometric to Kerr (we cannot expect
a global isometry to exist without further global conditions). 
This is the problem we want to solve in this paper.
More specifically, in section 2 we recall
the definition of the Simon tensor on the manifold of trajectories and
write down the main result proven in \cite{Simon}.
In section 3, we obtain the spacetime counterpart of the Simon tensor
in the strictly stationary region and show that it remains
regular at points where the Killing vector
becomes null. This requires some analysis
of the algebraic properties of the Simon tensor.
Then, we define a tensor
on any spacetime with a Killing vector (with no assumptions on 
its norm) such that it coincides with the
Simon tensor when projected down into the manifold of trajectories in the
strictly stationary region, and we obtain the consequences
of the vanishing of this tensor in terms of the Weyl tensor.
In section 4, we prove that
an asymptotically flat vacuum spacetime with a Killing  field which
is timelike near infinity and such that the corresponding
spacetime Simon tensor vanishes, is locally isometric to Kerr at every
point, thus extending the results in \cite{Simon}.

It is worth emphasizing 
that obtaining a spacetime characterization of Kerr is interesting
not only in order to understand better what is so special about Kerr,
but also for more practical purposes. The problem we have in mind is
the black hole uniqueness theorem. The existing proofs for
this theorem require rather strong hypotheses on the spacetime, like
connectedness  of the black hole event horizon (see Weinstein
\cite{We1}, \cite{We2} for interesting progress in the non-connected case),
non-existence of closed timelike curves in the domain of outer communication
(see Carter \cite{Carter3} for a discussion) and 
analyticity of the metric and the event horizon (see
Chru\'sciel \cite{Ch1}, \cite{Ch3}). This last point is crucial 
in order to apply the so-called Hawking rigidity theorem \cite{HE},
which ensures
the existence of a second Killing field  when the black hole
is rotating. It is reasonable to believe that the theorem still holds when
some (or perhaps all) of these
hypotheses are significantly relaxed. However, trying to prove
those results is a hard problem, and it is conceivable that
a more detailed knowledge of the Kerr geometry, and in particular
having at hand a spacetime characterization of Kerr
can prove helpful for attacking those questions.
In this respect, let us mention
that the characterization of Schwarzschild in terms of the conformal
flatness of the  hypersurfaces orthogonal to the static
Killing vector is an essential tool 
in all known proofs of the black hole uniqueness theorems in the
non-rotating case.

\section{The Simon tensor on the manifold of trajectories}

Let us start by fixing our definitions and conventions. 
Throughout this paper a $C^n$ spacetime denotes a paracompact,
Hausdorff, connected, $C^{n+1}$ four-dimensional manifold endowed with a $C^n$
Lorentzian metric $g$ of signature $(-1,1,1,1)$. Smooth means
$C^{\infty}$. We will also assume
that spacetimes are orientable and time-orientable.  The Levi-Civita
covariant derivative of $g$ is denoted by $\nabla$, the
volume form is $\eta_{\alpha\beta\gamma\mu}$ and our
sign conventions of the Riemann and Ricci tensors follow \cite{KSMH}.

Throughout this section $(\M,g)$ denotes a 
$C^2$ spacetime satisfying Einstein's vacuum field equations
$R_{\alpha\beta}=0$ and admitting a Killing vector field $\vec{\xi}$
which is timelike everywhere.
The construction of the Simon tensor is as follows. First, an 
equivalence relation $\approx$ is introduced
in $\M$ so that two points $p,q \in \M$ are
equivalent if and only if they belong to the same integral line
of $\vec{\xi}$.
If the spacetime satisfies the
chronology condition (see \cite{HE} for the definition) and the Killing
field is complete, then 
the quotient set $\N = \M /\approx$
is a differentiable manifold \cite{mantraj} such that the canonical
projection $\pi : \M \rightarrow \N$ is differentiable.
Then, there  exists a one-to-one correspondence between tensors
on $\N$ and tensors on $\M$ which are completely orthogonal to $\vec{\xi}$ 
(i.e. orthogonal to $\vec{\xi}$ with respect to any
index) and with vanishing Lie derivative along $\vec{\xi}$.

Then, the so-called norm and twist of the Killing field are defined
by $\lambda= - \xi^{\alpha} \xi_{\alpha}$ and
$\omega_{\alpha} = 
\eta_{\alpha\beta\gamma\delta} \xi^{\beta} \nabla^{\gamma}
\xi^{\delta}$. Since $\lambda >0$ in $\M$, the
tensors $h_{\alpha\beta} = g_{\alpha\beta} - \lambda^{-1}
\xi_{\alpha}\xi_{\beta}$ and $\gamma_{\alpha\beta} = \lambda
h_{\alpha\beta}$ are well-defined on $\M$. The corresponding tensors
on $\N$ are denoted by $\lambda$, $\omega_{i}$, $h_{ij}$ and
$\gamma_{ij}$ respectively (tensors on $\N$ carry Latin indices). 
Both $h_{ij}$ and $\gamma_{ij}$ are symmetric, non-degenerate and
positive definite. Denoting by $D$ the Levi-Civita covariant
derivative of $(\N, \gamma )$ and introducing the complex one form
\begin{eqnarray}
\sigma_{j} = D_{j} \lambda - i ~ \omega_{j},
\label{sigmai} 
\end{eqnarray}
the Simon tensor associated with $\vec{\xi}$ is defined\footnote{This
definition differs by a non-zero
factor from the original one in \cite{Simon}.} via \cite{Simon} 
\begin{eqnarray*}
S_{ijk} = 2 \left (\er_{\left [ k \right.} D_{\left . j \right ]}
\er_{i} - u_{\left [ k \right .}  \gamma_{\left . j \right ] i }
\right ), \hspace{1cm} u_k = \gamma^{ij} \sigma_{\left [k\right.}
D_{\left . j \right ]} \sigma_i.
\end{eqnarray*}
For later convenience, let us introduce a tensor
$\z_{ij} = D_{j} \er_i$ (which is symmetric by virtue of the
vacuum field equations) so that the Simon tensor reads
\begin{eqnarray*}
S_{ijk} = \z_{ij} \er_k - \z_{ik} \er_j 
- \frac{1}{2} \gamma_{ij} \left (\z \er_k - \er^{l} \z_{lk} \right )
+ \frac{1}{2} \gamma_{ik} \left (\z \er_j - \er^{l} \z_{lj} \right), 
\end{eqnarray*}
where $\z = \gamma^{ij} \z_{ij}$ and indices are raised with
$\gamma^{ij}$. One of the main results proven
in \cite{Simon} is (the concept of asymptotically flat end is
defined later in section 4)
\begin{theorem}{(Simon 1984)}
Let $\left (\M,g \right)$ be a $C^2$ vacuum spacetime with
a timelike Killing field $\vec{\xi}$.
Assume that the spacetime contains an asymptotically flat
end $\M^{\infty}$. If the Simon tensor associated with
$\vec{\xi}$ vanishes identically, then there
exists an asymptotically flat open submanifold $\M_1 \subset \M^{\infty}$
which is isometrically diffeomorphic to an open submanifold of the
Kerr spacetime.
\end{theorem}
The key idea of the proof is to show that the set of multipole moments defined
in the asymptotic end $\M^{\infty}$ of $(\M,g)$ coincides with the
set of multipole moments of the Kerr spacetime (for a certain mass and
angular momentum). Then, a previous result by Beig and Simon
\cite{BS} stating that two asymptotically flat
spacetimes with the same set of multipole
moments must be isometric in a neighbourhood of infinity completes the
proof.

\section{The spacetime Simon tensor}

In this section we will obtain
a tensor defined on any spacetime with a Killing vector,
which coincides with the Simon tensor when projected down into
to the manifold of trajectories (whenever this exists) and such that
it vanishes identically for the Kerr spacetime. 

Throughout this section $\left (\M,g \right )$ denotes a
$C^2$ vacuum spacetime admitting a non-trivial
Killing field $\vec{\xi}$. We do not require that the orbits
of this Killing vector are complete (i.e. the Killing vector $\vec{\xi}$
need not generate a one parameter isometry group).
A key object in our analysis is the exact two-form
${\bm F}_{\alpha \beta} = \nabla_{\alpha} \xi_{\beta}$
(we use boldface characters to denote two-forms). A complex two-form
$\cm{B}$ satisfying
$\cm{B}^{\star} = - i \cm{B}$, where
$\star$ is the Hodge dual operator, is called self-dual
(calligraphic letters will be used to denote them).
Given any real two-form
$\bm B_{\alpha\beta}$, its self-dual part $\cm{B}_{\alpha\beta}$, is
defined by  $\cm{B}_{\alpha\beta} = \bm{B}_{\alpha\beta} + i
\, \, \bm{B}^{\star}_{\alpha \beta}$. In particular,
\begin{eqnarray}
\cm{F}_{\alpha\beta} = \bm{F}_{\alpha\beta} + i \bm{F}^{\star}_{\alpha\beta}
\label{Fdual}
\end{eqnarray} 
is called {\it Killing form} throughout this paper.
Since most of the calculations in this section
involve two-forms, let us
recall some well-known identities (see e.g. \cite{Israel}). Let
$\bm{X}$ and $\bm{Y}$ be arbitrary two-forms on $\M$, then
\begin{eqnarray*}
\bm X_{\mu\sigma} \bm Y_{\nu}^{\ssigma} -  {\bm X^{\star}}_{\mu\sigma}
{\bm Y^{\star}}_{\nu}^{\ssigma} = \frac{1}{2} g_{\mu\nu} \bm
X_{\alpha\beta} \bm Y^{\alpha\beta} , \quad \bm X_{\mu\sigma}  {\bm
X^{\star}}_{\nu}^{\ssigma} =\frac{1}{4} g_{\mu\nu} \bm X_{\alpha\beta}
{\bm X^{\star}}^{\alpha\beta},
\end{eqnarray*}
which specialized to arbitrary self-dual two-forms,
$\bm{\cal X}$, $\bm{\cal Y}$, give
\begin{eqnarray}
\bm{\cal X}_{\mu\sigma} \bm{\cal Y}_{\nu}^{\ssigma} + \bm{\cal
Y}_{\mu\sigma} \bm{\cal X}_{\nu}^{\ssigma}  = \frac{1}{2} g_{\mu\nu}
\bm{\cal X}_{\alpha\beta} \bm{\cal Y}^{\alpha\beta}, \quad
{\bm{\cal
X}}_{\mu\sigma} \bm{\cal X}_{\nu}^{\ssigma} = \frac{1}{4} g_{\mu\nu}
\bm{\cal X}_{\alpha\beta} \bm{\cal X}^{\alpha\beta}.
\label{dual}
\end{eqnarray}
For any self-dual two-form $\bm{\cal Z}$ we have 
\begin{eqnarray}
\bm{\cal Z}_{\mu\sigma} \bm Z_{\nu}^{\ssigma} - \bm{\cal
Z}_{\nu\sigma} \bm Z_{\mu}^{\ssigma} =0,
\label{orto}
\end{eqnarray}
where $\bm{Z}$ denotes the real part of $\bm{\cal Z}$. 
From the equation
$\nabla_{\alpha} \nabla_{\beta} \xi_{\mu} = \xi^{\sigma} \,
C_{\sigma\alpha\beta\mu}$, which follows from the
Killing equations in vacuum, we obtain 
\begin{eqnarray}
\nabla_{\alpha} \F_{\beta\gamma} = \xi^{\sigma} \C_{\sigma\alpha\beta\gamma},
\label{df}
\end{eqnarray}
where $\C_{\alpha\beta\gamma\delta}$ is the so-called right self-dual
Weyl tensor, defined as
$\C_{\alpha\beta\gamma\delta} = C_{\alpha\beta\gamma\delta} +
\frac{i}{2} \eta_{\gamma\delta\rho\sigma}
C_{\alpha\beta}^{\,\,\,\,\,\,\,\, \rho\sigma}$.
This tensor shares all the
symmetries with the Weyl tensor, i.e.  it is a symmetric double two-form,
with vanishing trace, which satisfies the Bianchi identity
$\C_{\alpha \left [\beta\gamma\delta \right ] } =0$. It follows from
 (\ref{df}) that $\F$ satisfies Maxwell's equations
$d \F = 0$. The Ernst one-form is defined by 
\begin{eqnarray}
\er_{\mu} \equiv  2 \xi^{\alpha} \F_{\alpha\mu} = 
\nabla_{\mu} \lambda - i \omega_{\mu},
\label{Ernst}
\end{eqnarray}
which has as an immediate consequence that
\begin{eqnarray}
\sigma_{\mu} \sigma^{\mu} = -\lambda \F_{\alpha\beta} \F^{\alpha\beta}.
\label{s2}
\end{eqnarray}
The Ernst one-form $\er_{\alpha}$ is closed,
$\nabla_{\left [\mu\right .} \er_{\left . \nu \right ]} = 2 \F_{\alpha
\left [ \nu \right .} \FF_{\left . \mu \right ]}^{\,\,\,\,\alpha} +
\xi^{\alpha} \nabla_{\alpha} \F_{\mu\nu} = 0,
$ where $d\F = 0$ was used in the first equality and 
(\ref{orto}) and (\ref{df}) in the second one.
Let us assume for the moment that the set $\M^{+}= \left \{
p \in \M ; \lambda|_p > 0 \right \}$ is non-empty and that
$\N^{+}= \M^{+} / \approx$ is a manifold, so that 
the Simon tensor can be defined along the lines described in the previous
section. We want to translate the Simon tensor into a tensor in $\M^{+}$
and analyze whether it can be extended to all of $\M$.
The spacetime counterpart of $\sigma_i$ in (\ref{sigmai}) is
the Ernst one-form $\sigma_{\mu}$. The symmetric tensor $\z_{ij}$ will
also have a spacetime counterpart in $\M^{+}$, which we denote by
$\z_{\mu\nu}$. As we shall see below, $\z_{\mu\nu}$ is singular
on the boundary of $\M^{+}$ (if non-empty).
In order to see that the Simon tensor
remains nevertheless regular, we need to analyze 
the structure of the Simon tensor. To that end, let us
consider an arbitrary point $p \in \M^{+}$ and define the following
linear map between two-index symmetric, covariant tensors at $p$ which are
completely orthogonal to $\vec{\xi}$ 
and three-index covariant tensors at $p$
\begin{eqnarray}
U(Z)_{\alpha\beta\gamma} = Z_{\alpha\beta} \er_{\gamma} -
Z_{\alpha\gamma} \er_{\beta} - \frac{1}{2} h_{\alpha\beta} \left ( Z
\er_{\gamma} - \er^{\mu} Z_{\mu\gamma} \right )+ \frac{1}{2}
h_{\alpha\gamma} \left ( Z \er_{\beta} - \er^{\mu} Z_{\mu\beta} \right
),
\label{defU}
\end{eqnarray}
where $Z = g^{\mu\nu} Z_{\mu\nu}$. 
The algebraic properties of this map 
are summarized in the
following lemma, which is proven by straightforward calculation.
\begin{lemma}
\label{proper}
At any point $p \in \M^{+}$ and for any symmetric
tensor $Z_{\alpha\beta} |_p$ completely orthogonal to $\vec{\xi}$,
the tensor $U_{\alpha\beta\gamma} = \left . U(Z)_{\alpha\beta\gamma}
\right |_p$ defined  by (\ref{defU}) satisfies
\begin{eqnarray*}
1) ~ U_{\alpha\beta\gamma} \mbox{ is completely orthogonal to } \vec{\xi},
\hspace{9mm} 2) ~ U_{\alpha\beta\gamma} = - U_{\alpha\gamma\beta},
\hspace{9mm}
3) ~ U^{\alpha}_{\,\,\,\,\,\alpha\beta} = 0, \\
4) ~ U_{\left [\alpha\beta\gamma\right ]} = 0, \hspace{9mm} 
5) ~ \er^{\alpha} \er^{\beta} \left ( U_{\alpha\beta\gamma}
\er_{\delta} - U_{\alpha\beta\delta} \er_{\gamma} \right ) + 
\er^{\mu} \er_{\mu} \er^{\alpha} U_{\alpha\gamma \delta} = 0.
\hspace{9mm}
\end{eqnarray*}
\end{lemma}
Properties $2)$, $3)$ and $4)$ are standard for Cotton-like
tensors. Property $5)$ is specific for the Simon tensor.
A trivial calculation shows that the map 
$U(Z)$ satisfies
\begin{eqnarray}
U \left ( Z_{\alpha\beta} + s_1 h_{\alpha\beta} + s_2
\er_{\alpha} \er_{\beta} \right ) = U \left ( Z_{\alpha\beta} \right )
\label{kern1}
\end{eqnarray}
where $s_1$ and $s_2$ are arbitrary complex constants. This indicates
that the kernel of $U(Z)$ is at least two dimensional. The following
lemma shows that at points where $\F_{\alpha\beta}
\F^{\alpha\beta} \neq 0$ the kernel of $U(Z)$ is indeed  two-dimensional
and that $U(Z)$ is surjective onto
the set of tensors satisfying properties $1)$ to $5)$ above.
\begin{lemma}
Let $U_{\alpha\beta\gamma}$ be a tensor at $p \in \M^{+}$
satisfying properties $1)$ to $5)$ in lemma \ref{proper} and assume
that $\F_{\alpha\beta} \F^{\alpha\beta} |_p \neq 0$. Then 
$\sigma_{\mu} \sigma^{\mu} |_p \neq 0$ and
the general solution of the algebraic equation
$U(Z^{0} )_{\alpha\beta\gamma} = U_{\alpha\beta\gamma}$ is given by
\begin{eqnarray}
Z^{0}_{\alpha\beta} = \left . \frac{\sigma^{\gamma} 
\left ( U_{\alpha\beta\gamma} +U_{\beta\alpha\gamma} \right )
}{2\sigma_{\mu} \sigma^{\mu}} +
\frac{3
\sigma^{\nu} \sigma^{\gamma} \left (
U_{\nu\alpha\gamma}\sigma_{\beta} 
+U_{\nu\beta\gamma}\sigma_{\alpha}  
\right )} {2\left (\sigma^{\mu}\sigma^{\mu} \right )^2} 
+ s_1 h_{\alpha\beta}
+ s_2 \sigma_{\alpha} \sigma_{\beta} \right |_p
\label{aleqsol}
\end{eqnarray}
where $s_1$ and $s_2$ are arbitrary constants.
\label{inv}
\end{lemma}
{\it Proof.} Since $\lambda \F_{\alpha\beta} \F^{\alpha\beta} |_p \neq 0$,
equation (\ref{s2}) shows $\sigma_{\mu} \sigma^{\mu} |_p \neq 0$. 
Making use of
property (\ref{kern1}) the problem can be restricted to the one of finding
the general solution of
\begin{eqnarray}
\left .
{\tilde{Z}}_{\alpha\beta} \er_{\gamma} -
{\tilde{Z}}_{\alpha\gamma} \er_{\beta} + \frac{1}{2} h_{\alpha\beta} 
\er^{\mu} {\tilde{Z}}_{\mu\gamma} - \frac{1}{2}
h_{\alpha\gamma}  \er^{\mu} {\tilde{Z}}_{\mu\beta} \right |_p= U_{\alpha\beta\gamma},
\label{aleq}
\end{eqnarray}
for tensors ${\tilde{Z}}_{\alpha\beta}$ satisfying 
${\tilde Z}= {\tilde{Z}}_{\alpha\beta} \sigma^{\alpha}
\sigma^{\beta} |_p =0$. Contracting  (\ref{aleq}) with 
$\sigma^{\gamma}$ and $\sigma^{\alpha}$ alternatively,
it is easy to see that the only possible solution
of this equation
is given by the right-hand side of (\ref{aleqsol})
with $s_1 = s_2=0$. It remains to show that 
(\ref{aleq}) is fulfilled. This can be proven by introducing
three mutually orthogonal, unit complex vectors $e^i_{\mu}$, $i=1,2,3$,
at $p$ satisfying $\xi^{\mu} e^i_{\mu} |_p = 0$ and such
that $\sigma_{\mu} |_p =
(\sigma_{\beta} \sigma^{\beta})^{1/2} |_p e^3_{\mu}$.
Expanding $U_{\alpha\beta\gamma}$ in terms of these vectors it is
easy to obtain the
most general form allowed by the algebraic properties 1)-5)
in lemma \ref{proper}. Expanding also ${\tilde Z}_{\alpha\beta}$ in terms
of this basis,  it is a matter of simple calculation
to check that (\ref{aleq})  holds identically.
$\hfill \Box$

Let us return to the tensor $\z_{\alpha\beta}$.
The following lemma is key to show that the Simon tensor can be
extended to all of $\M$.
\begin{lemma}
$\z_{\nu\mu}$, i.e. the spacetime counterpart of $D_m \er_n$ on $\M^{+}$,
can be written as
\begin{eqnarray*}
\z_{\nu\mu} =2 \xi^{\alpha} \xi^{\beta} \C_{\alpha\mu\beta\nu} -
\frac{1}{2} \Fsq h_{\mu\nu} - \frac{1}{2 \lambda} \er_{\mu} \er_{\nu},
\end{eqnarray*}
where $\Fsq \equiv \F_{\alpha\beta} \F^{\alpha\beta}$.
\label{imp}
\end{lemma}
{\it Proof.} We shall start with the tensor $\z_{mn}$, defined on
$\N^{+}$ and find its spacetime counterpart.
Since the two metrics $h_{ij}$ and
$\gamma_{ij}$ are conformally related, it follows that
\begin{eqnarray*}
D_{m} \er_l = D^{h}_{m} \er_l + \frac{1}{2 \lambda } \left
 (h_{lm} \er^{k} D_k \lambda - \er_l D_m \lambda - \er_m D_l \lambda
 \right ),
\end{eqnarray*}
where $D^h$ denotes the Levi-Civita covariant derivative of $h$,
and indices are raised with $h^{ij}$. The spacetime
counterpart of the right-hand side can be written down
by recalling that each covariant derivative on $\N^{+}$ must 
be transformed into a covariant derivative on $\M^{+}$ and then
projected with $h^{\,\,\alpha}_{\beta} = \delta^{\,\,\alpha}_{\beta}
+ \lambda^{-1} \xi_{\beta} \xi^{\alpha}$. Hence
\begin{eqnarray*}
D_{m} \er_n \quad \longleftrightarrow \quad
\z_{\mu\nu}= h_{\mu}^{\,\,\mu^\prime} h_{\nu}^{\,\,\nu^\prime}
\nabla_{\mu^\prime}
\er_{\nu^\prime}  + \frac{1}{2 \lambda } \left (h_{\mu\nu}
\er^{\delta} \nabla_{\delta} \lambda - \er_{\mu} \nabla_{\nu} \lambda
- \er_{\nu} \nabla_{\mu} \lambda \right ),
\end{eqnarray*}
where the arrow stands for the one-to-one correspondence between
tensors on $\N^{+}$ and tensors on $\M^{+}$.
Using (\ref{df}) and (\ref{Ernst}), we immediately obtain
\begin{eqnarray*}
h_{\mu}^{\,\,\mu^\prime} h_{\nu}^{\,\,\nu^\prime} \nabla_{\mu^\prime}
\er_{\nu^\prime} =  2 h_{\mu}^{\,\,\mu^\prime} h_{\nu}^{\,\,\nu^\prime}
\FF_{\mu^\prime}^{\,\,\,\alpha} \F_{\alpha\nu \prime} + 
2 \xi^{\alpha} \xi^{\beta} \C_{\alpha\mu\beta\nu}.
\end{eqnarray*}
Thus, proving the lemma amounts to showing that
\begin{eqnarray}
2 h_{\mu}^{\,\,\mu^\prime} h_{\nu}^{\,\,\nu^\prime}  
\FF_{\mu^\prime}^{\,\,\,\alpha}
\F_{\alpha\nu^\prime} + \frac{1}{2 \lambda } \left (h_{\mu\nu}
\er^{\delta} \nabla_{\delta} \lambda - \er_{\mu} \nabla_{\nu} \lambda
- \er_{\nu} \nabla_{\mu} \lambda \right ) + \frac{1}{2} \Fsq
h_{\mu\nu} + \frac{1}{2 \lambda} \er_{\mu} \er_{\nu}= 0
\label{bigeq}
\end{eqnarray}
holds on $\M^{+}$.
The imaginary part of this equation is an immediate
consequence of  the imaginary part of (\ref{s2}) which reads
$\omega^{\alpha} \nabla_{\alpha} \lambda = \lambda \bm
F_{\alpha\beta} {\bm F^{\star}}^{\alpha\beta}$.
Regarding the real part of 
(\ref{bigeq}), the relation follows from the remarkable identity
(here comma stands for partial differentiation)
\begin{eqnarray}
\omega_{\mu} \omega_{\nu} + \lambda_{,\mu} \lambda_{,\nu} = \hspace{11cm}
\nonumber \\
\hspace{15mm}  2 \left ( \lambda
g_{\mu\nu} + \xi_{\mu} \xi_{\nu} \right ) \nabla^{\alpha} \xi^{\beta}
\nabla_{\alpha} \xi_{\beta} + g_{\mu\nu}  \lambda_{,\alpha} \lambda^{,
\alpha}  
- 4  \lambda \nabla_{\nu} \xi_{\alpha} \nabla_{\mu}
\xi^{\alpha}  - 4 \lambda^{, \alpha} \xi_{\left ( \nu \right .}
\nabla_{\left . \mu \right )} \xi_{\alpha}
\label{Ident}
\end{eqnarray}
which holds for any Killing vector. This identity can be proven,
after a somewhat long but trivial calculation, by expanding
$\eta_{\alpha\beta\gamma\delta} \,\eta_{\mu\nu\rho\sigma}$, appearing
in $\omega_{\mu}\omega_{\nu}$ on the left-hand side,  in terms of
products of $g_{\alpha\beta}$. $\hfill \Box$

This lemma shows that, although $\z_{\alpha\beta}$ becomes singular
at points where $\lambda$ goes to zero, the diverging part
belongs to the kernel of the map $U(\z)$. 
Therefore, the spacetime counterpart of the Simon tensor, which
is $S=U(\z)$, remains regular at the boundary of $\M^{+}$.
In other words, let us define the symmetric, trace-free tensor
\begin{eqnarray}
Y_{\mu\nu} = 2 \xi^{\alpha} \xi^{\beta} \C_{\alpha\mu\beta\nu} 
\label{Y}
\end{eqnarray}
which satisfies, due to (\ref{dual}), 
$\er^{\alpha} Y_{\alpha\beta} =
 - \lambda \xi^{\mu} \C_{\mu\beta\gamma\delta} \F^{\gamma\delta}$.
Using lemmas \ref{inv} and \ref{imp}, we find
that the spacetime counterpart of the Simon tensor at any
point $p \in \M^{+}$ reads
\begin{eqnarray*}
S_{\alpha\beta\nu} = U \left (Y \right )_{\alpha\beta\nu}
= Y_{\alpha\beta} \er_{\nu} -
Y_{\alpha\nu} \er_{\beta} - \frac{1}{2} \gamma_{\alpha\beta}
\xi^{\mu} \C_{\mu\nu\rho\delta} \F^{\rho\delta}
+\frac{1}{2} \gamma_{\alpha\nu} \xi^{\mu}
\C_{\mu\beta\rho\delta} \F^{\rho\delta}.
\end{eqnarray*}
All the objects in this expression are well-defined
in $\M$.
Moreover, this definition makes sense
irrespectively of whether the vacuum field equations hold or not.
Thus, let us put forward the following definition 
\begin{definition}
Let $(\M,g)$ be a $C^2$ spacetime with a Killing field 
$\vec{\xi}$. Construct $\F_{\mu\nu}$ and $\sigma_{\mu}$
through expressions (\ref{Fdual}) and (\ref{Ernst}).
The spacetime Simon tensor with respect to $\vec{\xi}$
is defined by
\begin{eqnarray*}
S_{\alpha\beta\nu} = U \left (Y \right )_{\alpha\beta\nu}
= 2 \xi^{\mu} \xi^{\rho} \C_{\mu\alpha\rho\beta} 
\er_{\nu} - 2 \xi^{\mu} \xi^{\rho} \C_{\mu\alpha\rho\nu} 
 \er_{\beta} - \frac{1}{2} \gamma_{\alpha\beta}
\xi^{\mu} \C_{\mu\nu\rho\delta} \F^{\rho\delta}
+\frac{1}{2} \gamma_{\alpha\nu} \xi^{\mu}
\C_{\mu\beta\rho\delta} \F^{\rho\delta},
\end{eqnarray*}
where $\gamma_{\alpha\beta} = \lambda g_{\mu\nu} + \xi_{\alpha}
\xi_{\beta}$ and $\lambda = - \xi^{\alpha} \xi_{\beta}$.
\end{definition}
Let us now consider the implications of the vanishing of the
spacetime Simon tensor. So, let $(V,g)$ be a spacetime with a Killing
field $\vec{\xi}$ such that the corresponding spacetime Simon tensor vanishes
everywhere. Consider a point $p \in V$ where $\vec{\xi}$ has
non-zero norm and $\Fsq$ is non-zero.
From lemma \ref{inv} and the fact that $Y_{\alpha\beta}$ is trace-free
it follows that
\begin{eqnarray}
\left .Y_{\alpha\beta} \right |_p  =
\left .2 \xi^{\mu} \xi^{\nu} \C_{\mu\alpha\nu\beta} \right |_p
= \left .\frac{Q(p)}{2} \left ( \sigma_{\alpha}\sigma_{\beta} + \frac{1}{3}
\gamma_{\alpha\beta} \Fsq \right ) \right |_p \label{elec}
\end{eqnarray}
where $Q(p)$ is an arbitrary complex constant. $\C_{\alpha\beta
\gamma\delta}$ is a double symmetric self-dual two-form which is
well-known to be uniquely characterized
by its electric and magnetic parts (see e.g. \cite{KSMH}).
Thus we have
\begin{eqnarray}
\left . \C_{\alpha\beta\gamma\delta} \right |_p =
Q(p) \left. \left (\F_{\alpha\beta} \F_{\gamma\delta} - \frac{1}{3} 
\I_{\alpha\beta\gamma\delta} \Fsq \right ) \right |_p
\label{weylSi}
\end{eqnarray}
where $\I_{\alpha\beta\gamma\delta} \equiv
(g_{\alpha\gamma}\, g_{\beta\delta}-g_{\alpha\delta}\,g_{\beta\gamma}
 + i \,\eta_{\alpha\beta\gamma\delta})/4$ is the metric in the
space of  self-dual
two-forms. The equality above holds because, by virtue of (\ref{elec}), 
the electric and magnetic parts of both sides coincide.
The following lemma will be important to prove the
characterization of Kerr.
\begin{lemma}
Let $(V,g)$ be a $C^3$ vacuum, not locally flat, spacetime with a
Killing vector
$\vec{\xi}$ which is non-null on a dense subset of $V$
and such that the corresponding spacetime Simon tensor vanishes everywhere. 
Assume also that $\Fsq \neq 0$ everywhere.
Then, the Ernst one-form is exact,(i.e. it exists a function
$\sigma$ such that $\sigma_{\alpha} = \nabla_{\alpha}
\sigma$) and the Weyl tensor and $\Fsq$ take the form
\begin{eqnarray*}
\C_{\alpha\beta\gamma\delta}  =
\frac{-6}{c - \sigma} 
\left (\F_{\alpha\beta} \F_{\gamma\delta} - \frac{1}{3} 
\I_{\alpha\beta\gamma\delta} \Fsq \right ), \hspace{1cm}
\Fsq = A \left (c -\sigma \right )^4
\end{eqnarray*}
where $c$ and $A\neq 0$ are complex constants.
\label{formWeyl}
\end{lemma}
{\it Proof.}  A trivial 
consequence (\ref{df}) and (\ref{weylSi}) is
\begin{eqnarray}
\nabla_{\alpha} \Fsq = \frac{2}{3} Q \Fsq  \sigma_{\alpha}.
\label{eqF2}
\end{eqnarray}
Since the spacetime is $C^3$ we can use the second Bianchi identities
which, in
terms of the right self-dual Weyl tensor, read
$\nabla_{\alpha} \C^{\alpha}_{\,\,\,\,\beta\gamma\delta}=0$. 
A straightforward calculation shows
\begin{eqnarray}
0 = 4 \xi^{\beta}\xi^{\gamma}
\nabla_{\alpha} \C^{\alpha}_{\,\,\,\,\beta\gamma\delta}=  
\frac{\lambda}{9} \Fsq Q^2 \sigma_{\delta}  - \frac{\Fsq}{3}
\left ( \lambda \nabla_{\delta} Q + 
\xi_{\delta} \xi^{\alpha} \nabla_{\alpha} Q \right )
- \sigma_{\delta} \sigma^{\alpha} \nabla_{\alpha} Q,
\label{expre}
\end{eqnarray}
which, after contraction with $\sigma^{\delta}$ yields
$\sigma^{\alpha} \nabla_{\alpha} Q = 
\frac{1}{6} \lambda \Fsq Q^2$,
where we used that $\lambda$ is zero at most on a set with empty
interior
and $\nabla_{\alpha} Q$ is continuous.  Inserting this expression back
into (\ref{expre}) we find $
\lambda \left (\nabla_{\delta} Q + \frac{1}{6} Q^2 \sigma_{\delta} \right )
- \xi_{\delta} \xi^{\alpha} \nabla_{\alpha} Q = 0$.
From $\C_{\alpha\beta\gamma\delta}
\F^{\alpha\beta} \F^{\gamma\delta} = 2/3 \left ( \Fsq \right )^2 Q$ we have 
$\xi^{\alpha} \nabla_{\alpha} Q =0$ and, therefore, 
\begin{eqnarray}
\nabla_{\alpha} Q + \frac{1}{6} Q^2 \sigma_{\alpha} = 0.
\label{EqQ}
\end{eqnarray}
The spacetime is not locally flat by assumption, so $Q$ is
not identically zero. Let us define the
set $K = \{p \in V ; Q(p)\neq 0 \}$ and suppose first that
$K \neq V$. Take a point $q \in\partial K$ (i.e.
in the topological boundary of $K$). Recall that the
Ernst one-form is closed on all of $V$ and therefore there exists
a sufficiently small open neighbourhood $V_q$ of $q$ where
the Ernst one-form is exact,
i.e. it exists a function $\sigma |_{V_q}$ such that
$\sigma |_{V_q} = \nabla_{\alpha} \sigma |_{V_q}$. On $K \cap
V_q$, equation (\ref{EqQ}) can be integrated to give
$\sigma - \frac{6}{Q} = c$ where $c$ is a complex constant. Since
$\sigma$ is well-defined in $V_q$ and in particular bounded at $q$
it follows that $Q$ cannot vanish on $q$,
against the assumption. Hence $Q$ is non-zero everywhere. Then, equation
(\ref{EqQ}) shows that $\sigma_{\alpha}$ is exact, i.e. $\sigma_{\alpha}
= \nabla_{\alpha} \sigma$. The connectedness of 
$V$ gives $Q = -6/(c - \sigma)$ everywhere. 
The integration of (\ref{eqF2}) completes the proof. $\hfill \Box$.

\section{The main Theorem}

A straightforward calculation shows that the spacetime Simon
tensor $S_{\alpha\beta\mu}$ 
with respect to the asymptotically timelike Killing vector vanishes
identically in the Kerr metric. The aim of this section is to prove
that the converse is also true in the asymptotically flat case. 
More precisely, we prove the following theorem
\begin{theorem}
Let $(V,g)$ be a smooth spacetime with the following
properties
\begin{enumerate}
\item The metric $g$ satisfies the Einstein vacuum field equations,
\item $(V,g)$ admits a smooth Killing field $\vec{\xi}$ such that the
spacetime Simon tensor associated to $\vec{\xi}$ vanishes everywhere, 
\item $(V,g)$ contains a stationary asymptotically flat four-end
$V^{\infty}$, $\vec{\xi}$ tends to a time translation at infinity
in $V^{\infty}$ and the Komar mass of $\vec{\xi}$ in $V^{\infty}$
is non-zero.
\end{enumerate}
Then $(V,g)$ is locally isometric to a Kerr spacetime.
\label{Main}
\end{theorem}
{\bf Remark 1}. By stationary asymptotically flat four-end we understand
an open submanifold $V^{\infty} \subset V$ diffeomorphic to
$I\times \left (\Bbb{R}^3 \setminus B(R) \right )$, ($I \in \Bbb{R}$
is an open interval and $B(R)$ is a closed ball of radius $R$), such that,
in the local coordinates defined by the diffeomorphism, the metric
satisfies
\begin{eqnarray*}
\left |g_{\mu\nu} - \eta_{\mu\nu} \right | +
\left |r \partial_i g_{\mu\nu} \right | \leq C r^{-\alpha},
\hspace{1cm} \partial_t g_{\mu\nu} =0
\end{eqnarray*}
where $C,\alpha$ are positive constants, $r = \sqrt{ \sum (x^i)^2}$ and
$\eta_{\mu\nu}$ is the Minkowski metric.  
Usually, the definition of asymptotically flat four-end requires 
$I = \Bbb{R}$, but this is not necessary for our purposes. 
A result by Kennefick and
\'{O} Murchadha \cite{KO} (see also Proposition 1.9 in \cite{Ch3})
shows that the Einstein field equations and the
existence of a timelike Killing vector
force $\alpha \geq 1$.
It is then well-known (see e.g. Beig and Simon \cite{BS1}) that
the metric can be brought into the asymptotic form
\begin{eqnarray}
g_{00} = -1 + \frac{2M}{r} + O(r^{-2}),
\hspace{6mm}
g_{0i} = - \epsilon_{ijk} \frac{4 S^j x^k}{r^3} + O(r^{-3}),
\hspace{6mm} g_{ij} = \delta_{ij} + O(r^{-1}), 
\label{AF}
\end{eqnarray}
where $M$ is the Komar mass \cite{Ko} 
of $\vec{\xi}$ in the asymptotically flat end $V^{\infty}$
and $\epsilon_{ijl}$ is the alternating Levi-Civita symbol.
It is worth noticing that assumption 3 in the theorem is used only in
order to prove
lemma \ref{exWeyl} below, which fixes the value of
the two arbitrary constants appearing in lemma \ref{formWeyl}.
Hence, assumption 3 in the theorem could be
replaced by any hypothesis under which lemma \ref{exWeyl} still holds.

{\bf Remark 2}. By Kerr spacetime we mean, as usual, the maximal analytic
extension of the Kerr metric, as described by
Boyer and Lindquist \cite{BY} and Carter \cite{Carter2}.
An element of the Kerr family will be denoted by $(V_{M,a},g_{M,a})$,
where $M$ denotes the Komar mass and
$a$ the specific angular momentum. In particular,  $(V_{M,0},g_{M,0})$ is 
the Kruskal extension of the Schwarzschild spacetime.

{\bf Remark 3}. 
As lemma \ref{formWeyl} above shows, the vanishing of the Simon tensor at 
points where $\Fsq \neq 0 $ and $\lambda \neq 0$ implies that the
Weyl tensor takes the form (\ref{weylSi}), which in particular shows that
the Petrov type of the Weyl tensor is D. The general solution for
vacuum spacetimes of Petrov type D was found by Kinnersley in 
\cite{Kinner}. Hence, we could in principle use his
results in order to prove the theorem and indeed we share with his
method the fact that we employ the Newman-Penrose formalism to prove
the theorem. However, we do not assume
analyticity of the metric as is usually done in the field
of exact solutions, where the main motivation is obtaining explicit
metrics satisfying Einstein field
equations. This requires extra care with the choice of tetrads
and coordinate systems and it is more convenient to produce a
self-contained proof. In particular, we try to use invariantly
defined quantities and avoid changing the null
tetrad in order to make some spin coefficients zero. By doing this, the
proof becomes geometrically more transparent and some insight is gained
into the path leading from the vanishing of the Simon tensor up to
the Kerr metric.

{\bf Remark 4}. The theorem states that for any point $p \in (V,g)$, there
exists an open neighbourhood $U_p$
of $p$ which is isometrically diffeomorphic to an open submanifold
of $V_{M,a}$. Since the characterization of Kerr in terms of the
spacetime Simon tensor is local and the only 
global requirement we make is the existence of an asymptotically flat
end (which, as pointed out in Remark 1, is only used in order to fix
the value of two constants), we should non expect in principle
that the local isometry
extends to an isometric embedding of $(V,g)$ into $(V_{M,a},g_{M,a})$. 
There can be topological obstructions for this
global embedding to exist. Analyzing this question in detail
would require classifying the spacetimes which are locally
isometric to Kerr and which are asymptotically flat (in the sense above, or
perhaps under the stronger requirement $I=\Bbb{R}$).
It would be necessary, among other
things, to determine the discrete isometry groups
of $(V_{M,a},g_{M,a})$ such that the quotient metric still
has an asymptotically flat four-end. This is not an easy problem because
the global structure of the Kerr spacetime is not particularly simple.
In the context of black hole
uniqueness theorems one is mainly interested in the domain of outer
communication. In this case, it is probably easy
to show, after assuming $I = \Bbb{R}$ in the definition
of asymptotically flat four-end, that the domain of outer
communication of $(V,g)$ is diffeomorphic to the domain of outer
communication of $(V_{M,a},g_{M,a})$. However, the analysis of
this problem will be relevant only if the
characterization of Kerr in terms of the spacetime Simon tensor
proves useful for extending the
black hole uniqueness theorems to the non-analytic case, and we will
not consider this question any further here.
We should emphasize, however,
that theorem \ref{Main} is ``semi-local'' 
because the existence of the local isometry is shown {\it everywhere}.

\hspace{3mm}

Throughout this section $(V,g)$ denotes a spacetime fulfilling the
requirements of theorem \ref{Main}. Let us normalize
the Killing vector and choose the integration constant in the twist potential
$\omega$ so that $\sigma \, \rightarrow 1$ at infinity in
$V^{\infty}$. A simple calculation using
the asymptotic form of the metric (\ref{AF})
in $V^{\infty}$ gives $\Fsq = - 4 M^2/r^4 + O (r^{-5})$. Since by
assumption $M\neq 0$ we have that the open submanifold
$\hat{V}_{\f} = \{ p \in V ; \Fsq |_p \neq 0 \}$ is
non-empty. Possibly after redefining $V^{\infty}$ as
an appropriate asymptotically flat open submanifold of $V^{\infty}$, we
can assume that $V^{\infty} \subset \hat{V_f}$.
Let us define the spacetime $(V_{f}, g_f)$
as the connected component of $\hat{V}_{f}$ containing the asymptotically
flat region $V^{\infty}$, with the induced metric. 
We have the following lemma
\begin{lemma}
Let $(V,g)$ satisfy the hypotheses of the theorem \ref{Main} and
$(V_f,g_f)$ be defined as above.
Then, the Weyl tensor and $\Fsq$ in $(V_{\f},g_{\f})$ take the form 
\begin{eqnarray*}
\C_{\alpha\beta\gamma\delta}  =
\frac{ - 6 }{1 -  \sigma} 
\left (\F_{\alpha\beta} \F_{\gamma\delta} - \frac{1}{3} 
\I_{\alpha\beta\gamma\delta} \Fsq \right ), \hspace{1cm}
\Fsq = - \frac{1}{4M^2}\left (1 -  \sigma \right )^4.
\end{eqnarray*}
\label{exWeyl}
\end{lemma}
{\it Proof.} In order to apply lemma \ref{formWeyl} we must 
show that the set of points where $\vec{\xi}$ has
non-zero norm is dense in  $V_{\f}$. We use the equation
\begin{eqnarray}
\nabla^{\mu} \nabla_{\mu} \sigma = - \Fsq = - 2 \left (
{\bm F}_{\alpha\beta} {\bm F}^{\alpha\beta} +
i \, {\bm F^\star}_{\alpha\beta} {\bm F}^{\alpha\beta} \right ),
\label{lapsigma}
\end{eqnarray}
which follows from
$\sigma_{\mu} = 2 \xi^{\alpha} \F_{\alpha\mu}$ and 
$\nabla^{\alpha} \F_{\alpha\beta}=0$. Assume there
exists an open set $U \subset V_{\f}$
where $\vec{\xi}$ has zero norm. The real part of (\ref{lapsigma}) implies
${\bm F}_{\alpha\beta}
{\bm F}^{\alpha\beta}|_{U} = 0$ and  
identity (\ref{Ident}) provides $\omega_{\mu} |_{U}=0$.
Using (\ref{lapsigma}) again we find 
$\Fsq |_U = 0$, which is impossible in $V_f$. Thus, lemma
\ref{formWeyl} can be applied.
The asymptotic form of the metric (\ref{AF}) implies
$\sigma  =  1 - 4M/r + O(r^{-2}) $ in $V^{\infty}$, which combined with the
asymptotic behaviour of $\Fsq$ gives
$c =1$ and $A= -1/(4M^2)$. $\hfill \Box$

It is convenient
to define a function $P|_{V_f} = (1-\sigma)^{-1} |_{V_f}$.
Equations (\ref{s2}) and (\ref{lapsigma}) read, in terms of $P$,
\begin{eqnarray}
\nabla_{\mu} P \nabla^{\mu} P = \frac{\lambda}{4M^2},
\hspace{1cm}
\nabla^{\mu} \nabla_{\mu} P = \frac{1}{4 M^2} \frac{2 \ov{P} -1 }{P \ov{P}},
\label{P2}
\end{eqnarray} 
where the bar denotes complex conjugate. Define the real functions
$y$ and $z$ by $P= y + i \, z$. Notice that $P$ is nowhere zero on $V_f$ and
thus $y$ and $z$ cannot vanish simultaneously in $V_f$.
As we shall see, the scalar functions
$y$ and $z$ are very closely related with the radial coordinate $r$ and 
the angular coordinate $\theta$ in the Boyer-Lindquist coordinates
of the Kerr metric. They have an intrinsic definition in terms
of the Ernst potential and will be essential for proving the existence
of the local isometry with the Kerr spacetime.
Since $\F_{\alpha\beta}$ is
self-dual and  $\Fsq |_{V_{\f}} = -1/(4M^ 2 P^4)$,
there exist two real, smooth, non-zero, null
vector fields $\vec{l}_{\pm}$ satisfying $\left( \vec{l}_{+}, \vec{l}_{-}
\right ) = -1$ such that 
\begin{eqnarray*}
\F_{\alpha\beta} = \frac{1}{4 M P^2} \left (
-l_{+\alpha} l_{-\beta}+l_{+\beta} l_{-\alpha} - i ~ \eta_{\alpha\beta
\gamma\delta} l^{\gamma}_{+}  l^{\delta}_{-} \right ),
\end{eqnarray*}
($\vec{l}_{\pm}$ are the eigenvectors $\F_{\alpha\beta}$, i.e.
$\F_{\alpha\beta} l_{\pm}^{\alpha} \propto l_{\pm\beta}$). The contraction
of this equation with $\xi^{\alpha}$ gives, after splitting into
the real and imaginary parts, 
\begin{eqnarray}
\nabla_{\beta} y = \frac{1}{2M} \left [ - \left (\vec{\xi},\vec{l}_{+} \right )
l_{-\beta} +\left (\vec{\xi},\vec{l}_{-} \right ) l_{+\beta} \right ]
\hspace{1cm}
\nabla_{\beta} z = - \frac{1} {2M}
\,\, \eta_{\alpha\beta\gamma\delta} \xi^{\alpha}
l_{+}^{\gamma} l_{-}^{\delta},
\label{split}
\end{eqnarray}
where the parentheses denote the scalar product with respect to the
metric $g_f$. Moreover ${\cal L}_{\vec{\xi}} \F_{\alpha\beta}=0$ provides,
after using $(\vec{l}_{+},\vec{l}_{-})=-1$, the relation
\begin{eqnarray}
\left [\vec{\xi}, \vec{l}_{\pm} \right ] = \pm \, C_1 \vec{l}_{\pm}
\hspace{5mm} \Longrightarrow \hspace{5mm} 
\xi^{\alpha} \nabla_{\alpha}
\left (\vec{\xi},\vec{l}_{\pm} \right )= \pm \, C_1 \left (\vec{\xi},
\vec{l}_{\pm} \right )
\label{commutator}
\end{eqnarray}
for some function $C_1$. Contracting $\sigma_{\beta} = 2 \xi^{\alpha}
\F_{\alpha\beta}$ with $\F_{\mu}^{\,\,\,\,\beta}$ and using
(\ref{split}) we get
\begin{eqnarray}
\xi_{\beta} = - 
\left (\vec{\xi},\vec{l}_{+} \right )
l_{-\beta} - \left (\vec{\xi},\vec{l}_{-} \right ) l_{+\beta} 
- 2M \, \eta_{\beta\mu\gamma\delta} \nabla^{\mu} z
\, l_{+}^{\gamma} l_{-}^{\delta}.
\label{Killing}
\end{eqnarray}
\begin{proposition}
The norms of $\nabla_{\alpha} z$ and $\nabla_{\alpha} y$ are
\begin{eqnarray}
\left . \nabla_{\alpha} z \nabla^{\alpha} z \right |_{V_{\f}} = \left .
\frac{B - z^2}{4 M^2 \left (y^2 + z^2 \right )} \right |_{V_{\f}} \hspace{1cm}
\left . \nabla_{\alpha}y \nabla^{\alpha} y \right |_{V_{\f}} =\left .
\frac{y^2 - y + B}{4 M^2 \left (y^2 + z^2 \right )} \right |_{V_{\f}} 
\end{eqnarray}
where $B$ is a non-negative constant. Moreover $z^2 |_{V_{\f}} \leq B$.
\label{normyz}
\end{proposition}
{\it Proof.} 
The real part of the first equation in (\ref{P2}) reads
\begin{eqnarray}
\nabla_{\alpha} y \nabla^{\alpha} y - \nabla_{\alpha} z \nabla^{\alpha} z
=  \frac{1}{4 M^2} \left (1- \frac{y}{y^2+ z^2} \right )
\label{reldydz}
\end{eqnarray}
so that we only need to prove $4 M^2 \nabla_{\alpha} z \nabla^{\alpha} z = 
(B- z^2)/(y^2+ z^2)$. Let us define a function 
$H = 4 M^2 P \ov{P} \nabla_{\alpha}
z \nabla^{\alpha} z$. 
Take an arbitrary point $p \in V_f$ and
consider a sufficiently
small open neighbourhood $U_p \subset V_f$ of $p$ so that
there exists a smooth complex vector field $\vec{m}|_{U_p}$
such that $\{\vec{l}_{+}, \vec{l}_{-},
\vec{m}, \vec{\ov{m}} \}$ 
is a null tetrad in $U_p$ with positive orientation, ( i.e. satisfying
$\eta_{\alpha\beta\gamma\delta} l_{+}^{\alpha} l_{-}^{\beta} m^{\gamma}
\ov{m}^{\delta} = - i$). We shall use the Newman-Penrose (NP) notation
to denote the Ricci
rotation coefficients associated with this null basis. We shall follow
the conventions in \cite{KSMH} except that we use $\vec{l}_{+}$ and
$\vec{l}_{-}$ instead of $\vec{k}$ and $\vec{l}$. 
Lemma \ref{exWeyl} shows that $\vec{l}_{\pm} $ are principal null
directions of the Weyl tensor and, from
the Goldberg-Sachs theorem \cite{GS}, 
they define geodesic and shearfree null congruences, which in NP
notation means
$\kappa = \sigma = \nu = \lambda = 0$.\footnote{The
spin coefficient $\sigma$ has nothing to do with the Ernst potential
$\sigma$ we have been using throughout. This is the only place
where it occurs and therefore no confusion should arise.} The subset
of NP equations we shall require are
\begin{eqnarray}
D \tau  = \rho \left (\tau + \ov{\pi} \right ) + \tau
\left (\epsilon - \ov{\epsilon} \right ), \hspace{4mm} &  & \hspace{4mm}
\delta \rho  = \rho \left (\ov{\alpha} + \beta \right ) + \tau \left  (
\rho - \ov{\rho} \right ), 
\label{seteqs1} \\
\Delta \pi  =  - \mu \left (\pi + \ov{\tau} \right ) + \pi
\left (\ov{\gamma} - \gamma \right ),
\hspace{4mm}  & & \hspace{4mm}
\ov{\delta} \mu = - \pi \left (\mu - \ov{\mu} \right ) - \mu \left (
\alpha + \ov{\beta} \right ).
\label{seteqs2}
\end{eqnarray}
For later use, let us also quote the commutators between the tetrad vectors,
which read
\begin{eqnarray}
\Delta \, D - D \, \Delta & = & \left (\gamma + \ov{\gamma} \right ) D
+ \left (\epsilon + \ov{\epsilon} \right ) \Delta - 
\left (\tau + \ov{\pi} \right) \ov{\delta} - \left (\ov{\tau} + \pi \right )
\delta \label{DeltaD}, \\
\delta \,D - D \, \delta & =& \left (\ov{\alpha} + \beta - \ov{\pi} \right ) D
- \left ( \ov{\rho} + \epsilon - \ov{\epsilon} \right ) \delta 
\label{deltaD}, \\
\delta \, \Delta - \Delta \, \delta & =& 
\left (\tau - \ov{\alpha} - \beta \right )
\Delta + \left (\mu - \gamma + \ov{\gamma} \right ) \delta 
\label{deltaDelta}, \\
\ov{\delta} \, \delta - \delta \, \ov{\delta} & =&
\left (\ov{\mu} - \mu \right ) D + \left ( \ov{\rho} - \rho \right ) \Delta
- \left (\ov{\alpha} - \beta \right ) \ov{\delta} - \left (\ov{\beta} - 
\alpha  \right ) \delta.
\label{deltadeltabar}
\end{eqnarray}
From lemma \ref{exWeyl}  the only non-zero component of the Weyl
spinor in this tetrad is $\Psi_2= - 1/(8m^2 P^3)$.
The Bianchi identities in the NP
formalism are $D P = - \rho P$, $\Delta P = \mu P$,
$\delta P = - \tau P$ and $\ov{\delta} P = \pi P$. Thus, 
(\ref{split}) becomes
\begin{eqnarray}
\left . \nabla_{\beta} y \right  |_{U_p} = P \left (\rho l_{-\beta} -
\mu l_{+\beta} \right ),
\hspace{1cm}
\left . \nabla_{\beta} z \right |_{U_p} =  - i \,\, \pi P m_{\beta}
+ i \,\, \tau P \ov{m}_{\beta}  
\label{Nabyz}
\end{eqnarray}
which implies $\rho P = \ov{\rho P }$,
$\mu P = \ov{\mu P } $ and $\tau P = \ov{\pi P } $. 
In terms of the spin coefficients, the function $H$ reads
$H = 8 M^2 P^2 \ov{P}^2 \pi \ov{\pi} \geq 0$ and equation 
(\ref{reldydz}) becomes $H = 8 M^2 P^3 \ov{P} \rho \mu + y - y^2 - z^2$.
A straightforward calculation
using equations (\ref{seteqs1})-(\ref{seteqs2}) gives
$D H = 0,$ $\Delta H = 0$, $\delta H = -2 i\,\ov{\pi}
\ov{P} z$.
Equation (\ref{Nabyz}) implies $D z = \Delta z =0$ and
$\delta z  = i \,\ov{\pi} \ov{P}$, from which
the constancy of $H+ z^2$ in
$U_p$ follows trivially. Since this holds for any $p \in V_f$ and
$V_f$ is connected, we obtain
$H |_{V_f} = B - z^2 |_{V_f}\geq 0$ where $B$ is a non-negative
constant. $\hfill \Box$

Let us now define two open sets 
$V_{\pm} = \{ p \in V_{\f} \,\, ; \,\, (\vec{\xi}, \vec{l}_{\pm})
|_p  \neq 0 \}$ and introduce the null vector fields
\begin{eqnarray*}
\left . \vec{s}_{\pm} \,\, \right |_{V_{\pm}} = \left . \frac{2M}{
\left (\vec{\xi},\vec{l}_{\pm} \right )} \,\, \vec{l}_{\pm}
 \right |_{V_{\pm}} , \hspace{1cm}
\left . \vec{k}_{\pm} \,\, \right |_{V_{\pm}} = \left . \frac{
\left (\vec{\xi},\vec{l}_{\pm} \right )}{2M} \,\,
\vec{l}_{\mp} \right |_{V_{\pm}}
\end{eqnarray*}
which satisfy $\left . \left (\vec{s}_{\pm},\vec{\xi} \right )
\right |_{V_{\pm}}=2M$,
$\left .\left (\vec{s}_{\pm},\vec{k}_{\pm} \right ) \right |_{V_{\pm}}=-1 $.
In terms of $\vec{s}_{\pm}$ and $\vec{k}_{\pm}$,
equation (\ref{split}) becomes
\begin{eqnarray}
\left. \nabla_{\alpha} y \right |_{V_{\pm}}= 
\left . \mp \left ( k_{\pm\alpha} + W s_{\pm\alpha} \right ) \right
|_{V_{\pm}}=
\left . \mp \left ( k_{\pm\alpha} + \frac{-y^2+y-B}{2M^2(y^2+z^2)}
s_{\pm\alpha} \right ) \right |_{V_{\pm}}.
\label{newdy}
\end{eqnarray}
where we have defined
\begin{eqnarray}
\label{defW}
\left . W \right |_{V_f} \equiv \left .
- \frac{\left (\vec{\xi}, \vec{l}_{+} \right )
\cdot \left (\vec{\xi}, \vec{l}_{-} \right )}{4M^2} \right  |_{V_f} =
\left . \frac{-y^2 + y - B}{8M^2 (y^2+ z^2)} \right |_{V_f},
\end{eqnarray}
the equality being a consequence of proposition \ref{normyz}.

The case in which the Killing vector $\vec{\xi}$ is hypersurface
orthogonal requires a
separate treatment. This case 
is characterized by $\nabla_{\mu} z=0$ in some open
set ${\cal U}$. Then, the imaginary part of the second equation in (\ref{P2})
implies $z |_{\cal U}=0$. Proposition \ref{normyz} shows $B=0$ and
hence that $z=0$ everywhere. 
\begin{proposition}
Assume that the Killing vector $\vec{\xi}$ of $(V_f,g_f)$ is 
hypersurface orthogonal.
Then, at any point $p \in V_{+} \cup V_{-}$ there exists a
neighbourhood $U_p$ of $p$ which can be isometrically
embedded into the Kruskal-Schwarzschild spacetime.
\label{Schwar}
\end{proposition}
{\it Proof.}
For definiteness, we shall give the proof only
for $p \in V_{+}$. The proof
for $V_{-}$ requires only a few sign changes and will be omitted.
So, take a point $p \in V_{+}$ and a sufficiently
small open neighbourhood $U_p \subset V_{+}$ of $p$. We shall use the
same notation as
in the proof of proposition \ref{normyz}. Since $P=y$ is real, so are
$\mu$ and $\rho$. Furthermore $z=0$ implies $\pi = \tau = 0$. The 
commutators (\ref{DeltaD}) and (\ref{deltadeltabar}) show
that the two-planes spanned by
$\{\vec{l}_{+},\vec{l}_{-} \}$ and by
$\{\vec{m},\vec{\ov{m}} \}$ are surface-forming. 
Denote by $\{S^{\iota}_{l} \}$ and $\{S^{\iota}_{m} \}$ ($\iota$
is an index labeling each surface) the corresponding
families of surfaces.  Thus, $U_p$ (or some open subset thereof
containing $p$) is foliated by two families of mutually
orthogonal two-surfaces.
We want to show that the induced metric in $S^{\iota}_{m}$ is that
of a two-sphere. To that end, we use the Gauss equation
to obtain the curvature scalar of the
induced metric in $S^{\iota}_{m}$. Recall that the second
fundamental form ${\cal K}$ of a surface is defined by 
${\cal K}(\vec{X},\vec{Y})=\left (\nabla_{\vec{X}} \vec{Y} \right
)^{\bot}$, where $\vec{X}, \vec{Y}$ are tangent vectors to the surface
and $\bot$ denotes the orthogonal projection to the
surface. A simple calculation using the definition of the NP coefficients
and (\ref{split}) gives
\begin{eqnarray}
\left . {\cal K} (\vec{X}, \vec{Y} )^{\beta} \right |_{S^{\alpha}_{m}} =
\left . - \frac{\nabla^{\beta} y}{y} \, \, g_f
 \left (\vec{X},  \vec{Y} \right )
\right |_{S^{\iota}_{m}}.
\label{secform}
\end{eqnarray}
Using the Gauss
equation (see e.g. \cite{KN}) combined with lemma \ref{exWeyl}, 
we obtain $\hat{R}^{\iota} = 
1/(2M^2 y^2)|_{S^{\iota}_{m}}$, where
$\hat{R}^{\iota}$ is the Ricci scalar of the surface
$S^{\iota}_{m}$. Since
$\delta y = 0$ (i.e. $y$ is constant on $S^{\iota}_{m}$) it follows that
$S^{\iota}_{m}$ has positive constant curvature and therefore
its metric is locally the round metric of a two-sphere of radius $r=2My$.
As $\nabla_{\alpha} y$ is nowhere zero on $V_{+} \cup V_{-}$, there
exist three functions $x^0, x^2, x^3$ defined on $U_p$ 
so that $\{x^0,y,x^2,x^3\}$ is a coordinate system on $U_p$ adapted
to the foliation, i.e  such
that $S^{\iota}_{m}$ is defined by $x^0 = \mbox{const}$, $y=\mbox{const}$
and $S^{\iota}_{l}$ is defined by $x^A= \mbox{const}, \: A=2,3$. Expression
(\ref{Killing}) shows that  
$\vec{\xi}$ is non-zero on $V_{+} \cup V_{-}$ and
tangent to $S^{\iota}_{l}$, hence
$\vec{\xi}\, |_{U_p}= \xi^0 \partial_{x^0}
|_{U_p}$ for some non-zero function $\xi^0$. The Killing
equations imply $\partial_{x^B} \xi^0 |_{U_p} = 0$. Regarding 
$\vec{s}_{+}$, (\ref{newdy}) implies $\vec{s}_{+}|_{U_p} =
\partial_y + s^0 \partial_{x^0} |_{U_p}$ for some function $s^0$. The
commutator (\ref{deltaD}) shows $\partial_{x^B} s^0 = 0$. Using
$[\vec{\xi},\vec{s}_{+} ]=0$ it is easy to see that there exists 
a function $u |_{U_p}$ such that $\{u,y,x^2,x^3\}$ is a coordinate
system in $U_p$ in which 
$\vec{\xi}= \partial_{u} |_{U_p}$ and $\vec{s}_{+}=\partial_{y} |_{U_p}$. 
It remains to determine $g_{AB} \equiv (\partial_{x^A}, 
\partial_{x^B} ) |_{U_p}$. 
Using equation (\ref{secform}) we easily obtain $g_{AB} =
4 M^2 y^2 g^{0}_{AB} (x^C)$,
for some functions $g^0_{AB}$ independent of $u$ and $y$. Imposing 
that the induced metric of $S^{\iota}_{m}$ is the metric of a sphere of
radius $r=2My$ we obtain, after defining $r = 2 M y$, that
\begin{eqnarray*}
\left . ds^2 \right |_{U_p} = \left . 
\left [ \left (-1 + \frac{2M}{r} \right ) du^2 + 
2 du \, dr + r^2 \left (d\theta^2 + \sin^2
\theta d \phi^2 \right ) \right ] \right |_{U_p},
\end{eqnarray*}
which proves that $U_p$ can be isometrically embedded into the
Kruskal-Schwarzschild spacetime. $\mbox{$\hfill \Box$}$

We now return to the general case (i.e. without imposing staticity).
Let us define the vector field
\begin{eqnarray}
\left .\vec{\eta} \, \right |_{V_{\f}} \equiv \left .
\frac{1}{2M} \left (B + y^2 \right) \vec{\xi}
+ \frac{\left (y^2 + z^2 \right)}{2M}
 \left [ \left (\vec{\xi}, \vec{l}_{-} \right ) \vec{l}_{+} 
+ \left (\vec{\xi}, \vec{l}_{+} \right ) \vec{l}_{-} \right ]
\right |_{V_{\f}},
\label{eta}
\end{eqnarray}
which, in $V_{\pm}$, can be rewritten as
\begin{eqnarray*}
\left . \vec{\eta} \, \right 
|_{V_{\pm}} = \left . \frac{1}{2M} \left (B + y^2 \right) \vec{\xi}
 + \left (y^2 + z^2 \right) \vec{k}_{\pm} +
\frac{y^2-y+B}{8M^2}  \vec{s}_{\pm} \,\,\, \right |_{V_{\pm}}.
\end{eqnarray*}
Let us still define another open set 
$V_1 = \{ p \in V_{\f} \, \, ; \,\, z^2 |p < B \}$ and a vector field
$b^{\alpha}  |_{V_1} = 
4M^2 (y^2 + z^2 )(B - z^2)^{-1} \nabla^{\alpha} z |_{V_1}$.
\begin{proposition}
Assume that the Killing vector $\vec{\xi}$ is not hypersurface
orthogonal. Then 
$\{ \vec{\xi}, \vec{s}_{\pm}, \vec{b}, \vec{\eta}\}$
define a holonomic basis of vector fields on $V_{\pm} \cap V_1$. Furthermore,
at any point $p$ in $(V_{+} \cup V_{-} ) \cap V_1$ there exists a
neighbourhood $U_p$ of $p$ that can be isometrically embedded into
the Kerr spacetime.
\label{basis}
\end{proposition}
{\it Proof.} The proof will again be given only for
$V_{+} \cap V_1$. The proof in $V_{-}\cap V_1$ is essentially the same.
First, we must prove that
$\{ \vec{\xi}, \vec{s}_{+}, \vec{b}, \vec{\eta}\}$ commute
and are linearly independent at any point on $V_{+} \cap V_1$.
An immediate consequence of (\ref{commutator}) is
$[\vec{\xi}, \vec{s}_{+}  ] |_{V_{+}} = \vec{0}$, 
$[\vec{\xi}, \vec{k}_{+}  ] |_{V_{+}} = \vec{0}$ and hence
$[\vec{\xi}, \vec{\eta} \, ]
|_{V_{+}} = \vec{0}$. $[\vec{\xi}, \vec{b} \, ] |_{V_1} = \vec{0}$ follows
easily
from the Killing equations.
Let us take an arbitrary point $p \in V_{+} \cap V_1$ and a sufficiently
small neighbourhood $U_p \subset V_{+} \cap V_1$ of $p$ where there
exists a complex null field $\vec{m}$ such that
$\{\vec{s}_{+},\vec{k}_{+},\vec{m},
\vec{\ov{m}} \}$ is a positively oriented null tetrad on $U_p$.  All the
relations obtained in the proof of proposition \ref{normyz} are still
valid when all NP spin coefficients referred to this new null tetrad.
Moreover, relation (\ref{newdy}) implies $DP=1$ and $\Delta P = W$
and therefore $\rho = -1/P$, $\mu= W/P$, which can be used to
simplify equations (\ref{seteqs1}) and commutators 
(\ref{DeltaD})-(\ref{deltadeltabar}). In particular, the second equation
in (\ref{seteqs1}) gives $\ov{\alpha} + \beta = \ov{\pi}$ which simplifies
the calculations considerably.
Using (\ref{Nabyz}), $\vec{b}$ takes the form
$ b^{\alpha} |_{U_p} = 4 \, i \, M^2 \left (y^2 + z^2 \right )(B - z^2)^{-1} 
( \ov{\pi} \ov{P} \ov{m}^{\beta} -   \pi P m^{\beta} )$.
$[\vec{s}_{+}, \vec{b} \, ]_{U_p} =
[\vec{\eta}, \vec{b} \, ]_{U_p} = \vec{0}$ follow after a simple
calculation using (\ref{seteqs1}), (\ref{deltaD}) and
(\ref{deltaDelta}). 
In order to prove $[\vec{s}_{+}, \vec{\eta} \,]|_{U_p}= \vec{0}$ we
need to evaluate $[\vec{s}_{+},\vec{k}_{+}]$. 
This is done by rewriting (\ref{Killing}) in the null tetrad, 
$\vec{\xi} \, |_{U_{p}}= 
2 M \left ( W \vec{s}_{+} - \vec{k}_{+} - P \pi \vec{m} - \ov{\pi}
\ov{P} \vec{\ov{m}} \right )|_{U_{p}}$. Imposing
$[\vec{\xi}, \vec{k}_{+} ] =\vec{0}$, we easily obtain
\begin{eqnarray*}
\left . \left [\vec{s}_{+} ,\vec{k}_{+} \right ] \right |_{V_{+}}=
\left . - \frac{2y}{y^2+z^2} \vec{k}_{+}
- \frac{y}{M \left (y^2 + z^2 \right)} \vec{\xi} + \frac{1-2y}{8M^2
\left (y^2+ z^2 \right )} \vec{s}_{+} \right |_{V_{+}}, 
\end{eqnarray*}
from which  $[\vec{s}_{+},\vec{\eta} ]|_{V_{+}}=\vec{0}$ is proven
by simple calculation. In order to show
that $\vec{\xi}$, $\vec{s}_{+}$, $\vec{\eta}$, $\vec{b}$
are linearly independent, we evaluate their scalar
products on $V_{+} \cap V_1$, which give
\begin{eqnarray}
\left (\vec{\xi}, \vec{\xi} \,\right )
 = -1 + \frac{y}{y^2+z^2}, \quad
\left (\vec{s_{+}}, \vec{\xi} \,\right )
= 2 M, \quad 
\left (\vec{\xi}, \vec{b} \,\right ) =0, \quad
\left (\vec{\xi}, \vec{\eta}\, \right )
 = \frac{y \left (B- z^2 \right )}{2 M
\left (y^2+z^2 \right )}, \nonumber \\
\left (\vec{s}_{+}, \vec{s}_{+} \right ) =0,
\quad \left (\vec{s}_{+}, \vec{b} \,\right )
=0, \quad \left (\vec{s}_{+}, \vec{\eta} \,\right ) = B - z^2, \quad 
\left (\vec{b}, \vec{b} \,\right )
= \frac{4M^2 \left (y^2 + z^2 \right)}{B - z^2 }, 
\nonumber \\
\left (\vec{b}, \vec{\eta} \,\right )=0, \quad
\left (\vec{\eta}, \vec{\eta} \,\right )
 = \frac{B-z^2}{4 M^2 \left (y^2+z^2 \right )}
\left [ z^2 \left (y^2-y+B \right ) + y^4+B y^2+By \right ].  
\label{scalar}
\end{eqnarray}
The determinant of the matrix formed by these scalar products (in
the obvious order) is
$-4M^2 (y^2+z^2) \neq 0$ which implies the linear independence of the
vectors. Let us define a positive constant $a$ by
$B = a^2/(4M^2)$ and introduce, as
before, a function $r$ by $r = 2 M y |_{V_{f}}$. Since
$z^2 |_{V_1} < B$, we can also define a function $\theta$ on $V_1$
by $z = a \cos \theta /(2M) |_{V_1}$. 
The holonomicity of 
$\{ \vec{\xi}, \vec{s}_{\pm}, \vec{b}, \vec{\eta}\}$ implies
that there exist
two functions $u|_{U_p}$ and $\phi |_{U_p}$ such
that $\{u,r,\theta,\phi \}$ is a coordinate system in $U_p$ satisfying
\begin{eqnarray*}
\left . \vec{\xi}\, \right |_{U_p} =\frac{\partial}{\partial u}, \quad
\left . \vec{s}_{+} \right |_{U_p} = 2 M \frac{\partial}{\partial r}, \quad
\left . \vec{b} \,\right |_{U_p} = - \frac{2 M}{a \sin \theta} 
\frac{\partial}{\partial \theta}, \quad
\left . \vec{\eta} \,\right |_{U_p} = \frac{a}{8M^3} 
\frac{\partial}{\partial \phi},
\end{eqnarray*}
which, after using (\ref{scalar}), yields
\begin{eqnarray*}
\left . ds^2 \right |_{U_p} = \left (-1 + \frac{2Mr}{r^2 + a^2 \cos^2 \theta}  
\right ) du^2 + 2 du \, dr + \frac{4 M a r \sin^2 \theta}{r^2 +
a^2 \cos^2 \theta} du \, d\phi + 2 a \sin^2 \theta dr \, d \phi  \\
\left . + \left (r^2 + a^2 \cos^2 \theta
\right ) d\theta^2 + \frac{\sin^2 \theta \left [
\left (r^2 + a^2 \right )^2 - a^2 \sin^2 \theta \left (r^2 - 2Mr +a^2 \right)
\right ] }{r^2 + a^2 \cos^2 \theta} d\phi^2 \right |_{U_p}.
\end{eqnarray*}
This is the form of the Kerr metric with mass $M$ and
specific angular momentum $a$ in the so-called Kerr coordinates.
This shows that $U_p$ is isometrically diffeomorphic to an open
submanifold of the Kerr spacetime. $\hfill \Box$

We have shown before that $\nabla_{\alpha} z |_U=0$
for an open set $U \subset V_f$ implies $z |_{V_{f}}=0$ and hence that
$\vec{\xi}$ is hypersurface orthogonal everywhere. This shows 
that $V_1$ is either dense in $V_f$ (when $\vec{\xi}$ is 
not hypersurface orthogonal)
or $V_1$ is empty  (when $\vec{\xi}$ is hypersurface orthogonal). The next
lemma gives
some properties of the complements of $V_1$ and $V_{+}\cup V_{-}$.
\begin{lemma}
The vector field $\vec{\eta}$ defined above 
is a Killing vector field in $V_f$. The complement
of $V_1$ in $V_f$ is 
$V_f \setminus V_1= \{ p \in V_f; \vec{\eta} \, |_p=\vec{0}\}$. Moreover,
\begin{itemize}
\item If $B=0$ then $V_{\f} \setminus (V_{+} \cup V_{-} ) = 
\{p \in V_f ; \vec{\xi} \, |_p= \vec{0}\}$
\item If $0 < B \leq 1/4$ then 
$V_{\f} \setminus (V_{+} \cup V_{-} ) = 
\{p \in V_f ; (\vec{\eta} - y_{+} \vec{\xi}\, ) \,|_p =\vec{0} \mbox{ or } 
(\vec{\eta} - y_{-} \vec{\xi}\,) \,|_p = \vec{0} \}$, where
$y_{\pm} = (1 \pm (1-4B)^{1/2})/4M$.
\item If $B>1/4$ then $V_{\f} \setminus (V_{+} \cup V_{-})$ is empty.
\end{itemize}
\label{fixedpoints}
\end{lemma}
{\it Proof.} 
In the particular case $B=0$ we have $z|_{V_f}=0$ and (\ref{eta}) implies
$\vec{\eta}|_{V_f}=\vec{0}$,
so that the first two claims of the lemma become obvious.
When $B\neq 0$, proposition \ref{basis} shows that $\vec{\eta}$ is a
Killing vector on the non-empty set
$(V_{+} \cup V_{-}) \cap V_1$,
which is linearly independent of $\vec{\xi}$. In the non-hypersurface
orthogonal
case $V_1$, is dense in $V_f$ and hence $\vec{\eta}$ is a Killing vector
in $\ov{V_{+} \cup V_{-}}$ (where the bar over a set denotes
its closure). Let us show that $\ov{V_{+} \cup V_{-}} = V_f$.
Assume that the open set $U_1 \equiv V_f \setminus (\ov{V_{+} \cup V_{-}})$
is non-empty. From the definition of $V_{\pm}$, we have
$(\vec{\xi}, \vec{l_{\pm}} )|_{U_1} = 0$ and (\ref{split}) implies
$\nabla_{\alpha} y |_{U_1}= 0$. The real
part of the second equation in (\ref{P2}) implies then $y |_{U_1} = 1/2$
and proposition \ref{normyz} fixes $B= 1/4$. The vector
field $\vec{\eta} - 1/(4M) \vec{\xi}$ is a non-zero Killing field in
$\ov{V_{+} \cup V_{-}}$ which vanishes identically on the open set $U_1$,
but this is impossible due to well-known properties of Killing vectors.
Thus, $\vec{\eta}$ is a Killing field everywhere. The second part
of the lemma is an obvious consequence of the definition of $\vec{\eta}$
(\ref{eta}). Regarding the statements for $V_f \setminus (V_{+} \cup V_{-})$,
the case $B=0$ is trivial from equation (\ref{Killing}), while
the last two statements follow easily from 
(\ref{defW}) and (\ref{Killing}), after noticing that  
the vanishing of $(\vec{\xi},\vec{l}_{+})$ implies that of $W$. $\hfill \Box$


We can now prove the main theorem.

{\it Proof of the Theorem.}
 
Propositions \ref{Schwar} and \ref{basis} show 
that the local isometry exists in a neighbourhood of any
point in $(V_{+} \cup V_{-})$ in the hypersurface orthogonal
case and of any point in 
$(V_{+} \cup V_{-}) \cap V_1$ in the non-hypersurface orthogonal
case. It remains to
show that the local isometry exists also in the complement of those
sets in $V_f$ which, from lemma \ref{fixedpoints} correspond to
the fixed points of certain Killing vectors in $V_f$.
Define the open set $W_1=  V_{+} \cup V_{-}$ in the hypersurface orthogonal
case
and $W_1=  (V_{+} \cup V_{-}) \cap V_1$ in the non-hypersurface orthogonal
one. Take an arbitrary point
$q \not\in W_1$ and a neighbourhood $U_q \subset V_f$ of
$q$. Since the
set of fixed points of a Killing vector is either an isolated point
or a smooth, two-dimensional, totally geodesic surface \cite{K},
the set $\tilde{U}_q = U_q \cap W_1$ is connected. Furthermore, any point
in $\tilde{U}_q$ has a neighbourhood which can be isometrically
embedded in the Kerr spacetime. Combining this two facts and
choosing $U_q$ small enough, it is easy to see that $\tilde{U}_q$
can be isometrically embedded into the Kerr spacetime.
This map can be extended to the closure of $\tilde{U}_q$ 
by continuity. Since $\tilde{U}_q$ is dense in $U_q$,
this defines a map $\Psi_q \, : \, U_q \,\, \rightarrow
(V_{M,a},g_{M,a})$. It is now a simple matter to show, using the
continuity of the metric, that $\Psi_q$ is an isometric embedding.

In order to complete the proof we need to show that $V_{\f} = V$.
Assume, on the contrary, that
$V_{\f}$ is a proper subset of $V$ and take a point
$q \in \partial V_{\f}$.
From continuity of $\Fsq$ we must have
$\lambda |_q =1$, hence $\vec{\xi}$ is timelike in some neighbourhood of
$q$.  Consider a smooth curve $\gamma_p (s)$ defined on some interval
$s_0 < s \leq 1$ such that $\gamma_p (s) \in V_f$, 
$\forall s \in (s_0,1)$,
$\gamma_p(1)=p$ and such that the tangent vector $\dot{\gamma} (s)$ is
orthogonal to $\vec{\xi}$ and of unit length (the existence of such a curve
is easy to establish).
Define the real function $Y(s) = (y \circ \gamma_p) (s)$.
Since $\sigma \rightarrow 1$ when we approach $p$ and $z$ remains bounded,
we must have $Y(s) \rightarrow \infty$ when $s \rightarrow 1$. 
Using $4 M \nabla_{\alpha} y \nabla^{\alpha} y |_{Y(s)} \rightarrow 1$ when
$s \rightarrow 1$, we can assume that $\nabla_{\alpha} y |_{Y(s)}$ is
spacelike for $s \in (s_0,1)$. Then 
\begin{eqnarray*}
\left (\frac{dY}{ds} (s) \right )^2 = \left (
\nabla_{\alpha} y |_{Y(s)} \dot{\gamma}^{\alpha}_p(s) \right )^2 
\leq \nabla_{\alpha} y \nabla^{\alpha} y |_{Y(s)} 
\end{eqnarray*}
where we have used the fact that $\{ \nabla_{\alpha} y |_{\gamma_p(s)},
\dot{\gamma}_p(s) \}$ define a spacelike two-plane  and we have
applied the Schwarz inequality. Hence $\frac{dY}{ds}$ stays bounded which
contradicts $Y \rightarrow \infty$ when $s \rightarrow 1$.
This completes the proof of the theorem. $\hfill \Box$.

Let us remark that the existence 
of a local isometry between two
spacetimes $(M_1,g_1)$ and $(M_2,g_2)$ on a dense subset of $M_1$
does not imply, in general, the existence of the local isometry
everywhere on $M_1$. It is easy to construct
counterexamples  when the manifolds are not analytic.
It is very likely that this cannot
happen when the target manifold is analytic and inextendible.
In any case, this result is not needed here because the
local isometry is ensured on an open, dense and {\it connected} set,
which makes the extension of the local isometry to all of $M_1$ simple.

\section{Discussion}

The characterization of the Kerr metric given by W.Simon \cite{Simon}
involved a three-index tensor in the quotient manifold which did not
have a clear geometrical interpretation (except in the hypersurface
orthogonal case, when
it is equivalent to the Cotton tensor). In this paper we have
translated the Simon tensor into the spacetime and have extended it
everywhere. By doing this, we can also obtain a geometrical
interpretation for the Simon tensor and hence a nice geometrical property
which characterizes the Kerr metric. Any spacetime with a non-trivial
Killing field $\vec{\xi}$ has a privileged two-form
${\bm F}_{\alpha\beta}= \nabla_{\alpha} \xi_{\beta}$, which can be
algebraically classified by analyzing the eigenvalue problem associated
with its self-dual two-form $\F_{\alpha\beta}$.
At points where $\F_{\alpha\beta}$ is so-called regular (i.e. $\Fsq \neq 0$),
there exist two principal
null directions defined as the eigenvectors of $\F_{\alpha\beta}$ 
(i.e. $\F_{\alpha\beta} l^{\beta} \propto l_{\alpha}$. At points
where $\F_{\alpha\beta}$ is singular (i.e $\Fsq =0$ with $\F_{\alpha\beta}
\neq 0$) there exists one principal null direction.
Similarly, the Petrov classification of the Weyl tensor
is the algebraic classification of the
endomorphism of the space of self-dual two-forms defined by 
\begin{eqnarray}
\C \left  ({\cm X} \right )_{\alpha\beta} = 
\C_{\alpha\beta}^{\,\,\,\,\,\,\,\,\mu\nu} {\cm X}_{\mu\nu}. 
\label{endom}
\end{eqnarray}
There exist four principal null directions
of the Weyl tensor (which  degenerate when the
Weyl tensor is algebraically special). A repeated
principal null direction of the Weyl tensor is a non-zero 
null vector field $\vec{l}$ satisfying
\begin{eqnarray*}
\C_{\alpha\beta\gamma\delta} l^{\beta} l^{\delta} = F l_{\alpha} l_{\gamma}
\end{eqnarray*}
for some function $F$. In section 3 we have shown that, at
points where $\vec{\xi}$ is non-null and $\F_{\alpha\beta}$ is
regular, the
vanishing of the Simon tensor is equivalent to 
the Weyl tensor taking the form (\ref{weylSi}). Let us state without
proof that this also holds at points $\Fsq =0$ provided
$\F_{\alpha\beta}\neq 0$. The Petrov classification of
(\ref{weylSi}) is very easy. In this case,
the eigenvalues of the endomorphism (\ref{endom}) are
$2/3 Q(p) \Fsq|_p$ (with eigenvector $\F_{\alpha\beta}|_p$)
and $-1/3 Q(p) \Fsq |_p$ (the eigenspace
of which is the set of self-dual two forms ${\cm X}$ satisfying
${\cm X}_{\alpha\beta} \F^{\alpha\beta} |_p = 0$). Hence, the
Weyl tensor (\ref{weylSi}) is of Petrov type 
D whenever $Q \Fsq |_p \neq 0$, type N when $\Fsq |_p =0, Q(p) \neq 0$
and Type 0 when $Q(p)=0$.  More importantly, the principal null directions
of the Weyl tensor coincide exactly with the principal null
directions of $\F_{\alpha\beta}$ (whenever $Q(p) \neq 0$).
At points where $\F_{\alpha\beta}$
is regular, each principal null direction of $\F_{\alpha\beta}$ is a
double principal null direction of the Weyl tensor
and at points where $\F_{\alpha\beta}$ is singular, the principal
null direction of $\F_{\alpha\beta}$ is a quadruple principal null
direction of the Weyl tensor. Hence, the Kerr metric is characterized
(among asymptotically flat vacuum solutions with a Killing field which
tends to a time translation at infinity) by the fact that
the principal null directions
of the Weyl tensor coincide with the principal null directions
of the Killing form $\F_{\alpha\beta}$.
This also shows why the Kerr metric enjoys such a privileged position 
among asymptotically flat stationary vacuum solutions.
The geometrical simplicity of this
characterization also indicates that the Kerr
metric is, in that sense, the simplest possible asymptotically flat
stationary vacuum metric.

A systematic study of the Killing
form $\F_{\alpha\beta}$ for electrovacuum
spacetimes admitting a Killing field has been carried out
recently by F.Fayos and C.F.Sopuerta \cite{PS}, who
have found a way to determine 
the principal null directions of $\F_{\alpha\beta}$ and of the
electromagnetic field. Several
conditions on the principal null directions can then be imposed,
thus restricting the possible geometries. For instance,
the condition that the principal null directions of $\F_{\alpha\beta}$
are aligned with the principal null directions of the electromagnetic
field can be imposed. Similarly, conditions like some (or all) principal
null directions of $\F_{\alpha\beta}$ being shear-free and/or geodesic
can also be studied. A particular interesting case would be studying
which electrovacuum spacetimes admitting a Killing vector field
are allowed such that the principal null directions of $\F_{\alpha\beta}$
and those of the electromagnetic field are geodesic and shear-free. From the
Goldberg-Sachs theorem, these null directions must then be repeated
principal null directions of the Weyl tensor, so that the Petrov 
type is D, N or 0 (depending on the degree of degeneracy). A straightforward
calculation shows that the Kerr-Newman metric has Petrov type D
with the two repeated principal null directions aligned with
the principal null directions of $\F_{\alpha\beta}$ (as in the
Kerr metric) {\it and} also aligned with the principal
null directions of the electromagnetic field.
It is a matter for future research to analyze 
whether the converse is also true
(after assuming asymptotic flatness) in order to 
obtain a characterization of the Kerr-Newman spacetime. 

\section*{Acknowledgements}

I am very grateful to Bobby Beig and Walter Simon for helpful discussions
and for a careful reading of the manuscript. I would also like to
thank Jos\'e M.M. Senovilla for comments on a previous version. 
I wish to thank the European Union for financial support
under the Marie Curie fellowship ERBFMBICT972520.

\end{document}